\documentclass{ws-procs9x6}

\begin{document}

\title{Determination of the fine structure constant with atom interferometry and Bloch oscillations}

\author{M. CADORET, E. de MIRANDES, P. CLAD\'E, C.~SCHWOB, F. NEZ, L. JULIEN $^*$ and F. BIRABEN}

\address{Laboratoire Kastler Brossel, ENS, CNRS, UPMC, 4 place Jussieu,\\ 75252 Paris CEDEX 05, France \\
  $^*$E-mail: biraben@spectro.jussieu.fr \\
  www.spectro.jussieu.fr}

\author{S. GUELLATI-KH\'ELIFA}

\address{Institut National de M\'etrologie, Conservatoire National des Arts et M\'etiers,\\ 61 rue Landy, 93210 La Plaine Saint Denis, France}

\begin{abstract}
  We use Bloch oscillations to transfer coherently many photon momenta to atoms. Then we can measure accurately the recoil velocity  $\hbar k/m$ and deduce the fine structure constant $\alpha$. The velocity variation due to Bloch oscillations is measured using atom interferometry. This method yields to a value of the fine structure constant $\alpha^{-1}= 137.035\,999\,45\,(62)$ with a relative uncertainty of about $4.5 \times 10^{-9}$.
\end{abstract}


\bodymatter

\section{Introduction}

\begin{figure}
\begin{center}
  \psfig{file=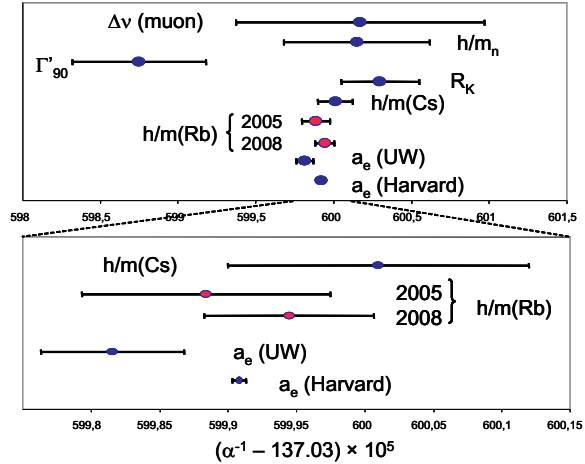,width=3.0in}
  \end{center}
  \caption{Determinations of the fine structure constant in different domains of physics. The most precise measurements are shown in the lower part of the figure. They are deduced from the anomaly of electron and from the ratio $h/m_{\rm Cs}$ and $h/m_{\rm Rb}$. We have taken into account the most recent result of the group of Gabrielse \cite{Gabrielse2008}. The two values deduced of $h/m_{\rm Rb}$ are presented in this paper.}
  \label{fig:fig:alpha2008}
\end{figure}

  The fine structure constant $\alpha$ is the fundamental physical constant characterizing the strength of the electromagnetic interaction. It is a dimensionless quantity, {\it{i.e.}}  independent of the system of units used. It is defined as:
\begin{equation}
\alpha=\frac{e^2}{4\pi\epsilon_0 \hbar c}
\end{equation}
where $\epsilon_0$ is the permittivity of vacuum, \emph{c} is the speed of light, \emph{e} is the electron charge and
\emph{$\hbar=h/2\pi$} is the reduced Planck constant. The fine structure constant is a key of the adjustment of the fundamental physical constants \cite{{codata02},{codata06}}. The different measurements of $\alpha$ are shown on Fig. \ref{fig:fig:alpha2008}. These values are obtained from experiments in different domains of physics, as the quantum Hall effect and Josephson effect in solid state physics, or the measurement of the muonium hyperfine structure in atomic physics. The most precise determinations of $\alpha$ are deduced from the measurements of the electron anomaly $a_e$ made in the eighties at the University of Washington \cite{VanDick} and, recently, at Harvard \cite{{Odom},{Gabrielse2007},{Gabrielse2008}}. This last experiment and an impressive improvement of QED calculations \cite{{Gabrielse},{Kino}} have lead to a new determination of $\alpha$ with a relative uncertainty of $3.7 \times 10^{-10}$. Nevertheless this last determination of $\alpha$ relies on very difficult QED calculations. To test it, other determinations of $\alpha$ are required, such as the values deduced from the measurements of $h/m_{\rm Cs}$ \cite{Wicht} and $h/m_{\rm Rb}$  ($m_{\rm Cs}$ and $m_{\rm Rb}$ are the mass of Cesium and Rubidium atoms) which are also indicated on the figure \ref{fig:fig:alpha2008}. In this paper, we present the measurements of $h/m_{\rm Rb}$ made in Paris in 2005 and 2008.

The principle of our experiment is the measurement of the recoil velocity $v_r$ of a Rubidium atom absorbing or emitting a photon ($v_r=\hbar k/m$ , where $k$ is the wave vector of the photon absorbed by the atom of mass $m$). As the relative atomic masses $A_r$ are measured very precisely, the measurement of $h/m_{\rm Rb}$ is a way to determine accurately $\alpha$  via the Rydberg constant $R_\infty$:
 \begin{equation}
\alpha^2=\frac{2R_\infty}{c}\frac{A_r({\rm Rb})}{A_r(e)}\frac{h}{m_{\rm Rb}}
\label{alpha-h/m}
\end{equation}
In this equation, the relative atomic mass of the electron $A_r(e)$ and the Rubidium ${A_r({\rm Rb})}$ are known with the relative uncertainties of $4.4\times 10^{-10}$ and $2.0\times 10^{-10}$ respectively \cite{{Beir},{Bradley}}. As the fractional uncertainty of $R_\infty$ is $7\times 10^{-12}$ \cite{{Udem},{Schwob}}, the factor limiting the accuracy of $\alpha$ is the ratio $h/m_{\rm Rb}$.

\section{Principle of the experiment}

The principle of the experiment is to coherently transfer as many recoils as possible to the atoms ({\it{i.e.}} to accelerate them) and to measure the final velocity distribution. In our experiment, the atoms are efficiently accelerated by means of $N$ Bloch oscillations (BO). The velocity selection and velocity measurement are done with Raman transitions.
The experiment develops in three steps. i) Firstly, we select from a cold atomic cloud of $^{87}{\rm Rb}$ a bunch of atoms with a very narrow velocity distribution. This selection is performed by a Doppler velocity sensitive counter-propagating Raman transition. In 2005, we have used a $\pi$-pulse to transfer the atoms from the $F=2$ to the $F=1$ hyperfine levels of $^{87}{\rm Rb}$. In 2008, we have modified the experimental scheme to take advantage of Ramsey spectroscopy: we use a pair of $\pi/2$ pulses which produces a fringe pattern in the velocity space. ii) Secondly, we transfer to these selected atoms as many recoils as possible by means of Bloch oscillations as explained later. iii) Finally, we measure the final velocity of the atoms by a second Raman transition which transfers the atoms from the $F=1$ to the $F=2$ hyperfine level. In short, we have used two different pulse sequences, the $\pi-\rm{BO} -\pi$ and $\pi/2-\pi/2-\rm{BO} -\pi/2-\pi/2$ configurations.

The accuracy of our measurement of the recoil velocity relies in the number of recoils ($2N$) that we are able to transfer to the atoms. Indeed, if we measure the final velocity with an accuracy of $\sigma_v$, the accuracy on the recoil velocity measurement $\sigma_{v_r}$ is:
 \begin{equation}
\sigma_{v_r}=\frac{\sigma_v}{2N}
\label{incertitude}
\end{equation}

\begin{figure}
\begin{center}
  \psfig{file=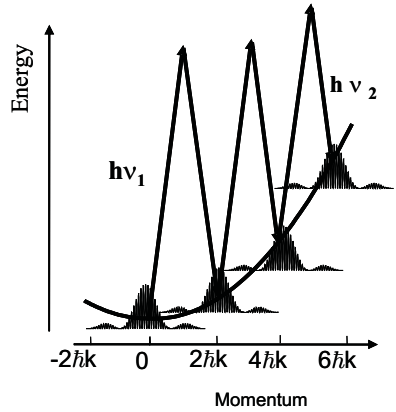,width=2.0in}
  \end{center}
    \caption{Acceleration of cold atoms with a frequency chirped
standing wave. The variation of energy versus momentum in the
laboratory frame is given by a parabola. The energy of the atoms
increases by the quantity $4(2j+1)E_{r}$ at each cycle. The Ramsey fringe patterns represents the momenta distribution of the atoms in the $F=1$ hyperfine level.}
  \label{fig:Parabole}
\end{figure}

Bloch oscillations have been first observed in atomic physics by the groups of Salomon in Paris and Raizen in Austin \cite{{Dahan},{Peik},{Raizen}}. In a simple way, Bloch oscillations can be seen as Raman transitions where the atom begins and ends in the same energy level, so that its internal state ($F=1$ for $^{87}{\rm Rb}$) is unchanged while its velocity has increased by $2v_r$ per Bloch oscillation. This is illustrated on Fig. \ref{fig:Parabole} which shows the atomic kinetic energy versus the atomic momentum. The velocity distribution obtained after the $\pi/2-\pi/2$ selection is also represented. Bloch oscillations are produced in a one dimensional optical lattice which is accelerated by linearly sweeping the relative frequency of two counter propagating laser beams (frequencies $\nu_1$ and $\nu_2$). The frequency difference $\Delta\nu$ is increased so that, because of the Doppler effect, the beams are periodically resonant with the same atoms ($\Delta\nu=4(2j+1)E_{r}/h$, $j=0,1,2,3..$ where $E_{r}/h$ is the recoil energy in frequency units and $j$ the number of transitions). This leads to a succession of rapid adiabatic passages between momentum states differing by $2\hbar k$. In the solid-state physics approach, this phenomenon is known as Bloch oscillations in the fundamental energy band of a periodic optical potential. The atoms are subject to a constant inertial force obtained by the introduction of the tunable frequency difference $\Delta\nu$ between the two waves that create the
optical potential \cite{Dahan}.

We now describe the acceleration process following the Bloch formalism. If, after the selection, the atom has a well defined momentum $\hbar q_0$ with $|q_0|<k$, the atomic wave function is modified when the optical potential is increased adiabatically (without acceleration) and becomes in the first energy band:
\begin{equation}
|\Psi_{0, q_0}\rangle=\sum_l{\phi_0(q_0+2lk)|q_0+2lk\rangle} \label{momentum space}
\end{equation}
with $l\in \mathbb{Z}$. Here $|q_0\rangle$ designs the ket associated to a plane wave of momentum
$q_0$ and the amplitudes $\phi_0$ correspond to the Wannier function  \cite{Wannier} in momentum space of the first band. When the potential depth is close to zero, the limit of the Wannier function $\phi_0$ is 1 over the first Brillouin zone and zero outside. On the contrary, if the potential depth is large, the Wannier function selects several components in the velocity space. When the optical lattice is accelerated adiabatically, the Wannier function is continuously shifted in the momentum space following the relation:
\begin{equation}
|\Psi(t)\rangle=\sum_{l}\phi_{0}(q_0 + 2lk - mv(t)/\hbar) |q_{0}+2lk\rangle\\
\label{wannier1}
\end{equation}
where $v(t)$ is the velocity of the optical lattice. The enveloping Wannier function $\phi_{0}$ is shifted by $mv(t)$ in momentum space. After the acceleration, the potential depth is decreased adiabatically and, in equation \ref{wannier1}, the Wannier function selects only one component of the velocity distribution. At the end, the wave function is $ |\Psi\rangle=|q_{0}+2Nk\rangle$. If $\Delta v$ is the velocity variation due to the acceleration, the number of Bloch oscillations $N$ is such as $|\hbar q_0+m\Delta v-2N\hbar k|<\hbar k$. Consequently, if the initial atomic velocity distribution fits the first Brillouin zone, it is exactly shifted by $2Nv_r$ without deformation, as it is shown, on Fig. \ref{fig:Parabole}, for the velocity distribution produced by a pair of $\pi/2$ pulses.

\section{Results in the $\pi-\rm{BO} -\pi$ configuration}

Our experimental setup has been previously described in detail \cite{{Clade1},{Clade2}}.  Briefly, we use a magneto-optical trap (MOT) and an optical molasses to cool the atoms to about $3~\mu K$. The determination of the velocity distribution is performed using a $\pi-\pi$ pulses sequence of two vertical counter-propagating laser beams (Raman beams) \cite{SENV}: the first pulse with a fixed frequency $\delta_{sel}$, transfers atoms from $5 S_{1/2}$, $\left|F=2, m_F = 0\right>$ state to $5 S_{1/2}$,
$\left|F=1,m_F = 0\right>$ state, into a narrow velocity class (width of about $v_r/15$). Then a laser beam resonant with the $5S_{1/2}~(F=2)$ to $5P_{1/2}~(F=3)$ cycling transition pushes away the atoms remaining in the ground state $F=2$. Atoms in the state $F=1$ make $N$ Bloch oscillations in a vertical accelerated optical lattice. We then perform the final velocity measurement using the second Raman $\pi$-pulse, whose frequency is $\delta_{meas}$. The populations of the $F=1$ and $F=2$ levels are measured separately by using a time of flight technique. To plot the final velocity distribution we repeat this procedure by scanning the Raman beam frequency $\delta_{meas}$ of the second pulse.

\begin{figure}
  \begin{center}
  \psfig{file=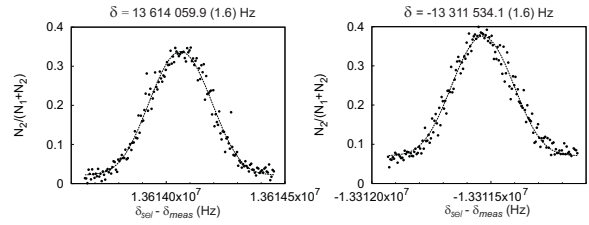,width=3.0in}
  \end{center}
    \caption{Velocity spectra obtained in the $\pi-\rm{BO} -\pi$ configuration. Here $N_1$ and $N_2$ are respectively the number of atoms in $F=1$ and $F=2$ after the acceleration process. These two spectra are obtained by performing the Bloch acceleration upwards or downwards. The frequency difference between these spectra corresponds 1780 recoil velocities.}
  \label{fig:Spectre1}
\end{figure}

To avoid spontaneous emission and to reduce other stray effects (light shifts and refraction index), the Raman lasers and the optical lattice are blue detuned by $\sim 1~\mathrm{THz}$ and $\sim 40~\mathrm{GHz}$ respectively from the one photon transition. The delay between the two $\pi$-pulses is $12$ ms and their duration $3.4$ ms. The optical potential depth is $70~E_r$. For an acceleration of about $2000~\mathrm{ms}^{-2}$ we transfer about $900$ recoil momenta in $3$~ms with an efficiency greater than $99.97\%$ per recoil. To avoid atoms from reaching the upper windows of the vacuum chamber, we use a double acceleration scheme: instead of selecting atoms at rest, we first accelerate them using Bloch oscillations and then we make the three steps sequence: selection-acceleration-measurement. This way, the atomic velocity at the measurement step is close to zero. In order to eliminate the effect of gravity, we make a differential measurement by accelerating the atoms in opposite directions (up and down trajectories) keeping the same delay between the selection and measurement $\pi$-pulses. The ratio $\hbar/m$ can then be deduced from the formula:
\begin{equation}
\frac{\hbar}{m}= \frac{(\delta_{sel}-\delta_{meas})^{up} -
(\delta_{sel}-\delta_{meas})^{down}}{2(N^{up}+N^{down})k_B
(k_1+k_2)} \label{eq:mynum}
\end{equation}
where $(\delta_{meas}-\delta_{sel})^{up/down}$ corresponds respectively to the center of the final velocity distribution for the up and the down trajectories, $N^{up/down}$ are the number of Bloch oscillations in both opposite directions, $k_B$ is the Bloch wave vector and $k_1$ and $k_2$ are the wave vectors of the Raman beams. Moreover, the contribution of some systematic effects (energy level shifts) is inverted when the direction of the Raman beams are exchanged: for each up or down trajectory, the Raman beams directions are reversed and we record two velocity spectra. Finally, each determination of  $h/m_{\rm Rb}$  and $\alpha$ is obtained from 4 velocity spectra. Fig. \ref{fig:Spectre1} shows two velocity spectra for the up and down trajectories.

 The  determinations of $ h/m_{\rm Rb}$ and $\alpha$ have been derived from 72 experimental data point taken during four days. In these measurements, the number of Bloch oscillations were $N^{up}=430$ and $N^{down}=460$. Then, the effective recoil number is $2(N^{up}+N^{down})$=1780. The dispersion of these n=72 measurements is $\chi^2/(n-1)=1.3$ and the resulting statistical relative
uncertainty on $h/m_{\rm Rb}$ is $8.8 \times 10^{-9}$. This corresponds to a relative statistical uncertainty on $\alpha$ of $4.4 \times 10^{-9}$. All systematic effects affecting the experimental measurement have been analyzed in detail in reference \cite{Clade2}. The total correction due to the systematic effects is $(10.98 \pm 10.0) \times 10^{-9}$ on the determination of $h/m_{\rm Rb}$. With this correction, we obtain for $\alpha$:
\begin{equation}
    \alpha^{-1}= 137.035\,998\,84\,(91) \quad [6.7\times 10^{-9}]
\end{equation}
This value of the fine structure constant is labeled $h/m({\rm Rb}) 2005$ on Fig. \ref{fig:fig:alpha2008}.

\begin{figure}
  \begin{center}
  \psfig{file=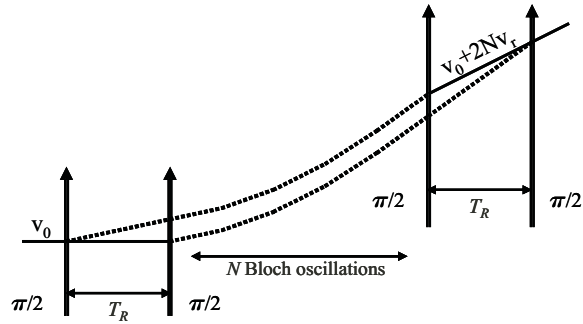,width=3.0in}
  \end{center}
    \caption{Scheme of the interferometer used for the measurement of $h/m_{\rm Rb}$. The first pair of $\pi/2$ pulses produces a fringe pattern in the velocity distribution which is measured by the second pair of $\pi/2$ pulses. Between these two pairs of pulses, the atoms are accelerated upwards or downwards. The solid line corresponds to the atom in the $F=2$ state, and the dashed line to the $F=1$ state.}
  \label{fig:Interferometre}
\end{figure}

\section{Measurement of the fine structure constant by atom interferometry}

We describe in this section the results obtained in the $\pi/2-\pi/2-\rm{BO} -\pi/2-\pi/2$ configuration. The scheme of this interferometric method is shown on Fig. \ref{fig:Interferometre}. The frequency resolution is now determined by the time $T_R$ within each pair of pulses while the duration of each $\pi/2$ pulse determines the spectral width of the pulses and the number of atoms which contribute to the signal. This interferometer is similar to the one of the reference \cite{Wicht}, except that effective Ramsey $k$-wavevectors point in the same direction. Consequently, this interferometer is not sensitive to the recoil energy, but only to the velocity variation due to Bloch oscillations which take place between the two sets of $\pi/2$ pulses.

\begin{figure}
  \begin{center}
  \psfig{file=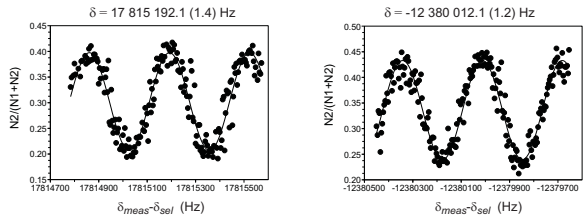,width=3.0in}
  \end{center}
    \caption{Velocity spectra obtained in the $\pi/2-\pi/2-\rm{BO} -\pi/2-\pi/2$ configuration. Here $N_1$ and $N_2$ are respectively the number of atoms in $F=1$ and $F=2$ after the acceleration process. The spectrum on the left corresponds to the downwards acceleration (600 Bloch oscillations) and on the right to the up acceleration (400 Bloch oscillations). The frequency difference between these spectra corresponds 2000 recoil velocities.}
  \label{fig:Spectre2}
\end{figure}

As in the $\pi-\rm{BO} -\pi$ configuration, a value of $h/m_{\rm Rb}$ is deduced from four spectra obtained with the upwards or downwards acceleration and by exchanging the directions of the Raman beams. An example of two spectra is shown on Fig. \ref{fig:Spectre2}. In this case, the total number of Bloch oscillations is $N^{up}+N^{down}=1000$, corresponding to $2000$ recoil velocities between the up and down trajectories. The duration of each $\pi/2$ pulse is $400 \mu$s and the time $T_R$ is $2.6$ ms (the total time of a pair of $\pi/2$ pulses is $3.4$ ms). For these experiments, the blue detuning of the Raman lasers is $310$ GHz. By comparison with the $\pi-\rm{BO} -\pi$ configuration (see Fig. \ref{fig:Spectre1}) the resolution is better: the half period of the fringes is about $160$ Hz when the line width of the spectra of Fig. \ref{fig:Spectre1} was about $500$ Hz. Nevertheless, the reduction of the uncertainties is not in the same ratio. This is due to the phase noise of the Raman laser which becomes more important. To lower this effect, we have set up an active anti vibration system: then the precision of each frequency measurement increases by about a factor of two. After this amelioration, to improve the resolution, it is tempting to increase the time $T_R$ between the $\pi/2$ pulses. Nevertheless, another limitation appears, which is the size of the vacuum cell. Indeed, when we make the selection, the velocity of the atoms is close to $2Nv_r$ and, during the selection, the atom travels a distance of $2Nv_rT_R$. For example, in the case of the spectra of Fig. \ref{fig:Spectre2}, the total displacement of the atoms is about $88$ mm and $53$ mm for the down and up trajectories (the upper window of the vacuum cell is $70$ mm above the center of the cell): practically, with the parameters corresponding to the spectra of Fig. \ref{fig:Spectre2}, we use all the size of the cell. Consequently, if we want to increase the delay $T_R$, we have to reduce the number $N$ of Bloch oscillations and there is no benefit.

\begin{figure}
  \begin{center}
  \psfig{file=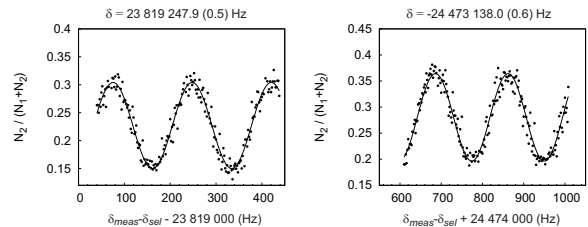,width=3.0in}
  \end{center}
    \caption{Velocity spectra obtained in the $\pi/2-\pi/2-\rm{BO} -\pi/2-\pi/2$ configuration with the atomic elevator.  The spectrum on the left corresponds to the downwards acceleration ($800$ Bloch oscillations) and on the right to the upwards acceleration ($800$ Bloch oscillations). The frequency difference between these spectra corresponds $3200$ recoil velocities.}
  \label{fig:Spectre3}
\end{figure}

To surpass this limit, we have developed a method, called \emph{atomic elevator}, to better use the volume of the vacuum cell. The idea is to move the atoms to the top or the bottom of the cell before making the sequence described above. Then, we use the total size of the cell to accelerate and decelerate the atoms. To displace the atoms, we accelerate the atoms with $300$ Bloch oscillations during $4$ ms and, after a dead time of $13$ ms, we decelerate the atoms with $300$ Bloch oscillations. This sequence displaces the atoms by about $50$ mm. With this technique, we have increased at the same time $N$ to $800$ and $T_R$ to $5.7$ ms. Fig. \ref{fig:Spectre3} shows two records obtained with these parameters. The visibility of the fringes is similar to the one of Fig. \ref{fig:Spectre2} and the half period of the fringe is about $90$ Hz. Now the frequency difference between the two spectra corresponds to $3200 v_r$. During the selection, the atomic velocity is about 10 m/s and the atom travels a distance close to 6 cm. This shows the limitation due to the size of the vacuum cell.

We present now the result deduced from $221$ measurements of $h/m_{\rm Rb}$. For these measurements, we have used the two methods described previously, with and without the atomic elevator. The total number of Bloch oscillations $N^{up}+N^{down}$ varies from $200$ to $1600$. The dispersion of these $n=221$ measurements is $\chi^2/(n-1)=1.85$ and the resulting relative statistical uncertainty on $\alpha$ is $3 \times 10^{-9}$. The systematic effects are similar to the ones described in reference \cite{Clade2}. The two main effects are due to the geometry of the laser beams and to the second order Zeeman effect. To evaluate the first effect, we have measured the wave front curvatures with a Shack-Hartmann wave front analyzer (HASO 128 from Imagine Optics). From these measurements, we have obtained a correction of $(-11.9 \pm 2.5) \times 10^{-9}$ on the determination of $\alpha$. As explained above, the effect of parasitic level shifts is eliminated by inverting the direction of the Raman beams. Nevertheless, this assumes that the measurements are made exactly at the same position when the direction of the Raman beam is inverted.  In fact, these positions are not exactly the same because the directions of the recoils given at the first Raman transition are opposite. For the timing used in our experiment, they differ by about  $\delta_z=300~\mu\mathrm m$. We have precisely measured the spatial magnetic field variations to control this effect. This correction depends of the number of Bloch oscillations. For example, in the case of the records of Fig. \ref{fig:Spectre3}, its value is $(7 \pm 1) \times 10^{-9}$. Finally the relative uncertainty due to the systematic effects is $3.4 \times 10^{-9}$ and we obtain for $\alpha$:
\begin{equation}
    \alpha^{-1}= 137.035\,999\,45\,(62) \quad [4.5\times 10^{-9}]
\end{equation}
This value corresponds to the point labeled $h/m({\rm Rb}) 2008$ on Fig. \ref{fig:fig:alpha2008} and is in agreement with our 2005 measurement. Our two results are also in agreement with the most precise value deduced from the electron anomaly (labeled $a_e$(Harvard) on Fig. \ref{fig:fig:alpha2008}).

\section{Conclusion}

We have presented two determinations of the fine structure constant $\alpha$. Depending on the Raman pulses arrangement ($\pi-\rm{BO} -\pi$ or $\pi/2-\pi/2-\rm{BO} -\pi/2-\pi/2$ configurations), our experiment can run as an atom interferometer or not.  The comparison of the two resulting values, which are in good agreement, provides an accurate test of these methods.
The comparison with the value extracted from the electron anomaly experiment \cite{Gabrielse2008} is either a strong test of QED calculations or, assuming these calculations exact, it gives a limit to test a possible internal structure of the electron. Our goal is now to reduce the relative uncertainty of $\alpha$. We are building a new experimental setup with a larger vacuum chamber. With the new cell, we plan to multiply the number of Bloch oscillations by a factor of three. Then, it will be possible to reduce the uncertainty at the $10^{-9}$ level to obtain an unprecedent test of the QED calculations.

\section*{Acknowledgments}

This experiment is supported in part by the Laboratoire National de M\'etrologie et d'Essais (Ex. Bureau National de
  M\'etrologie)(Contrat 033006), by IFRAF (Institut Francilien de Recherches sur les Atomes Froids) and by the Agence Nationale
  pour la Recherche, FISCOM Project-(ANR-06-BLAN-0192).

\end{document}